\documentclass[12pt]{article}
\usepackage{graphicx}
\usepackage{url}

\usepackage{amssymb}
\usepackage{amsmath}
\usepackage{latexsym}

\def\R{\mathbb{R}}
\def\s{{\sigma}}
\def\t{{\tau}}
\def\o{{\omega}}
\def\X{{\mathcal X}}
\def\K{{\mathcal K}}
\def\W{{\mathcal W}}

\newcommand{\figlab}[1]{\label{fig:#1}}
\newcommand{\figref}[1]{\ref{fig:#1}}

\title{Computational Geometry Column~45}
\author{%
Joseph O'Rourke\thanks{
Dept. of Computer Science, Smith Col\-lege, North\-ampton, 
MA 01063, USA.
orourke@\allowbreak cs.\allowbreak smith.\allowbreak edu.
Supported by NSF Distinguished Teaching Scholar Grant DUE-0123154.
}
}
\date{}
\begin{document}
\maketitle

\begin{abstract}
The algorithm of Edelsbrunner for surface reconstruction
by ``wrapping'' a set of points in $\R^3$ is described.
\end{abstract}

Curve reconstruction~\cite{o-cgc38-00} seeks to find a
``best'' curve passing through a given finite set of points,
usually in $\R^2$.
Surface reconstruction seeks to find a best surface passing through
a set of points in $\R^3$.
Both problems have numerous applications, usually deriving from
the need to reconstruct the curve or surface from a sample.
Both problems are highly underconstrained, for there are usually
many curves/surfaces through the points.
Surface reconstruction in particular is notoriously difficult to control.
Although significant advances have been made in recent 
years~\cite{d-csr-04}---especially in the direction of performance 
guarantees based on sample density---we turn here to a beautiful 
and now relatively old 
``wrapping'' algorithm due to Edelsbrunner,
which, although implemented in 1996 at 
Raindrop Geomagic,
has been published only recently~\cite{ede-srwfp-03}
after issuance of a patent in 2002.
\begin{figure}[htbp]
\begin{minipage}[t]{0.5\linewidth}
\centering
\includegraphics[width=\linewidth]{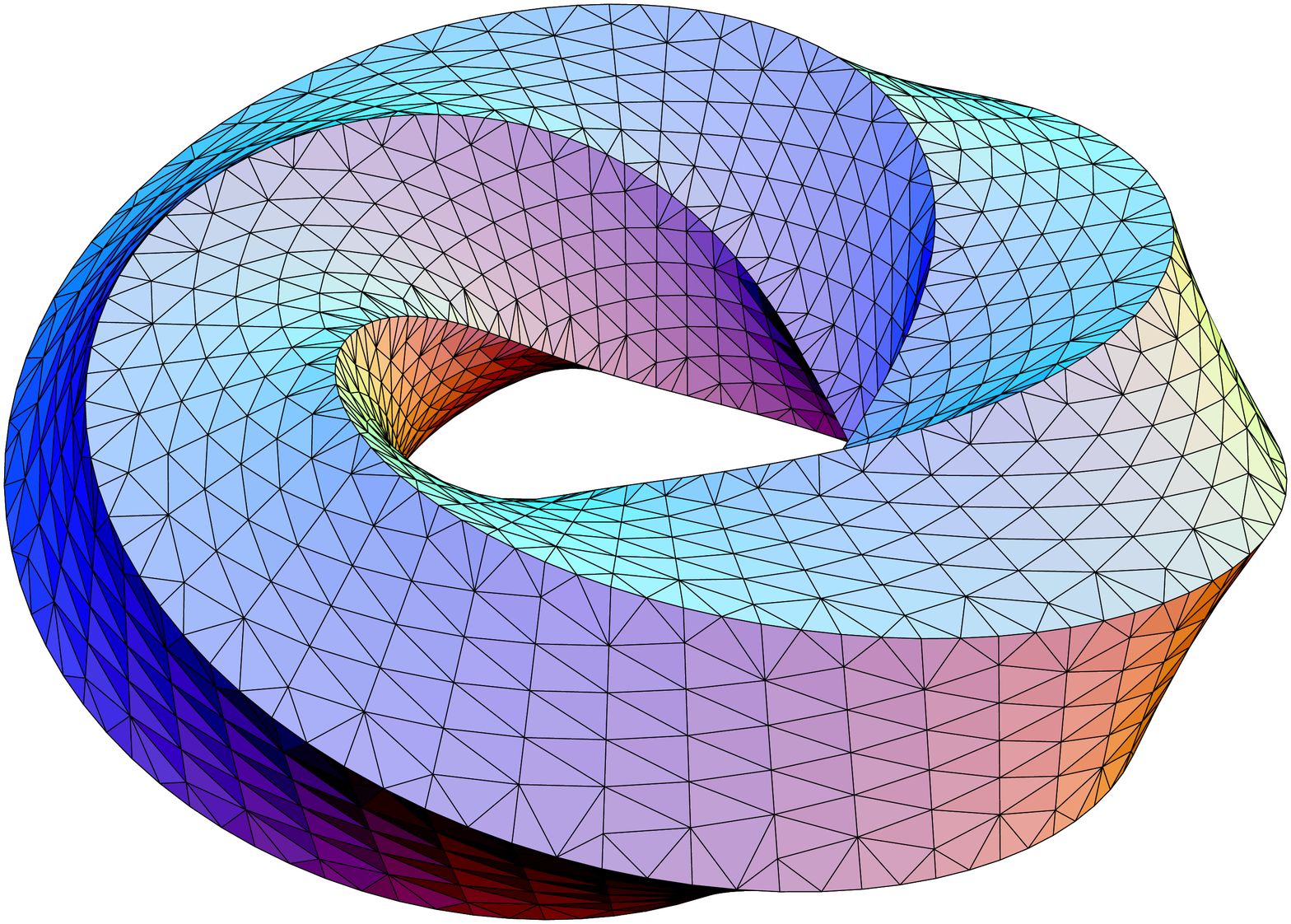}
\caption{Torus, pentagonal cross-section.}
\end{minipage}%
\begin{minipage}[t]{0.5\linewidth}
\centering
\includegraphics[width=\linewidth]{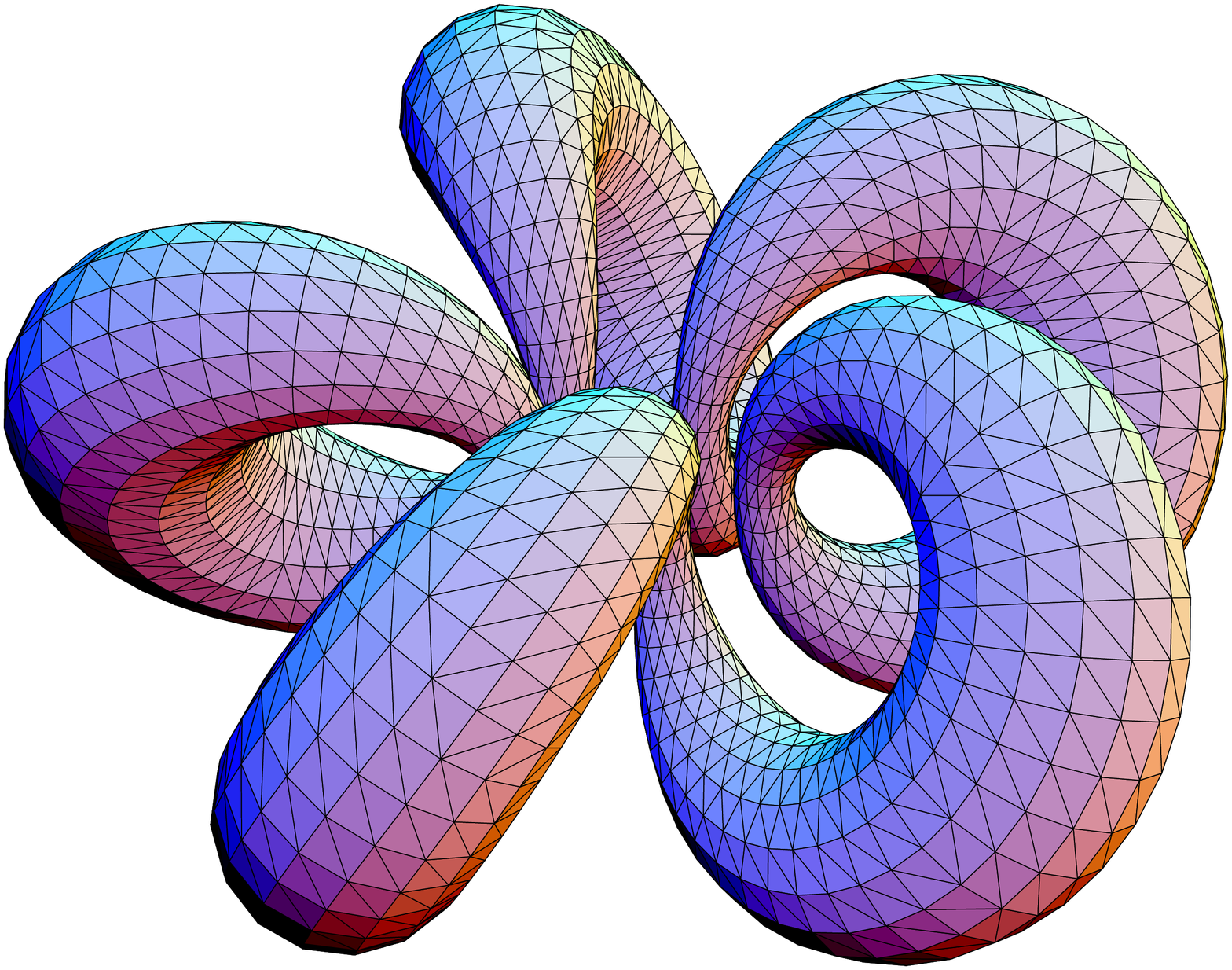}
\caption{Smooth, twisted torus.}
\figlab{tubes}
\end{minipage}
\end{figure}

Sample results of the algorithm are illustrated in Figs.~1 and ~2.%
\footnote{
  {\tt .stl} (stereolithography) files for shapes 
  from \url{http://www.cs.duke.edu/~edels/Tubes/}.
}
Although both of these examples reconstruct surfaces of genus one, 
we concentrate on the genus-zero case (a topological sphere) and only mention
extensions for higher genus reconstructions.

An attractive aspect of the algorithm is that it reconstructs a unique
surface without assumptions on sample density and without
adjustment of heuristic parameters.
Although the algorithm uses discrete methods, underneath it relies on
continuous Morse functions.
The discrete scaffolding on which the algorithm depends is
the Delaunay complex, which we now informally describe.
A \emph{simplex} is a point, segment, triangle, or tetrahedron.
A \emph{simplicial complex} $\K$ is a ``proper'' gluing together of
simplicies, in that (1)~if a simplex $\s$ is in $\K$, then so are all
its faces, and (2)~if two simplices $\s$ and $\s'$ are in $\K$, then
either $\s \cap \s'$ is empty or a face of each.
Let $S$ be the finite set of points whose surface is to be reconstructed.
The \emph{Delaunay complex} Del~$S$ 
is the dual of the Voronoi diagram
of $S$.  Under a general-position assumption, Del~$S$ contains a simplex 
that is the convex hull of the sites $T \subset S$ 
iff there is an empty
sphere that passes through the points of $T$.
The outer boundary of Del~$S$ is the convex hull of $S$.
Augmenting Del~$S$ with a dummy ``simplex'' $\o$ for the
space exterior to the hull, covers $\R^3$.

The algorithm seeks to find a ``wrapping'' surface $\W$,
a connected simplicial subcomplex in Del~$S$.
It accomplishes this by finding a simplicial subcomplex $\X$ of Del~$S$
whose boundary is $\W$.  The vertices of $\X$ will be precisely the
input points $S$, and the vertices of $\W$ will be a subset of $S$.

The algorithm uncovers $\X$ in Del~$S$ by ``sculpting'' 
away simplices from Del~$S$ one-by-one,
starting from $\o$, until $\X$ remains.
The simplices are removed according to an acyclic partial ordering.
It is the definition of this ordering that involves continuous
mathematics.

A function $g(x)$ assigns to every point $x \in \R^3$ a number dependent on the
closest Voronoi vertex.
In particular, if $x$ is in a tetrahedron $T$ of Del~$S$ whose
empty circumsphere has center $z$ and radius $r$, then 
$g(x) = r^2 - || z-x ||^2$.
Thus $g(x)$ is zero at the corners of $T$ and rises to $r^2$ at $z$,
the closest Voronoi vertex.
Points outside the hull are assigned an effectively infinite value.
$g(x)$ is continuous but not smooth enough to qualify as a Morse function,
needed for the subsequent development.  It will suffice here to claim
that $g$ can be smoothed sufficiently to define the vector field
$\nabla g$, and from this, by a limiting process, \emph{flow curves}
through every point $x \in \R^3$ aiming toward higher values.

These flow curves are in turn used to define an acyclic relation on
all the simplices of Del~$S$ and $\o$.
Let $\t$ and $\s$ be two simplices (of any dimension) and $v$ a
face shared between them.  For example, if $\t$ and $\s$ are both
tetrahedra, $v$ could be a triangle, or a segment, or a vertex.
Define the \emph{flow relation} ``$\to$'' so
that $\t \to v \to \s$ if there is a flow curve passing from
int~$\t$ to int~$v$ to int~$\s$.\footnote{
  int~$v$ is the interior of $v$;
  for a $v$ a vertex, int~$v = v$.
}

A sink of the relation is a simplex that has no flow successor.
$\o$ is always a sink (recall $g(x)$ is large outside
the hull), with the hull faces of Del~$S$ its immediate
predecessors.
Sinks are like critical points of the flow,
with the simplices that gravitate toward a sink corresponding to
a stable manifold in Morse terminology.

A key theorem is that the flow relation on simplices is acyclic,
which reflects the increase of $g(x)$ along every flow curve.
The algorithms starts with $\o$ and methodically ``collapses'' its
flow predecessors
until no more collapses are possible, yielding the complex $\X$.

Let $v$ be a face of $\t$; then $\t$ is called a \emph{coface} of $v$.%
\footnote{One can think of this is a \emph{co}ntaining \emph{face},
  although its origins are more in complementary topological terminology.}
Assume $\t \to v$; for example, $\t$ might be a tetrahedron and $v$
one of its edges, with the flow from $\t$ through $v$.
We give some indication of when the pair $(v,\t)$ is collapsible, without
defining it precisely.
First, $\t$ must be the highest dimension coface of $v$,
and $v$ should not have any cofaces not part of $\t$.
Thus, $v$ is in a sense ``exposed.''
Second, the flow curves should pass right through every point of $v$
(as opposed to running along or in $v$).
Collapse of the pair removes all the cofaces of $v$, thus eating away
the parts of $\t$ sharing $v$.

A second key theorem is that any sequence of collapses from $\o$
leads to the same simplicial complex $\X$.
Collapses also maintain the homotopy type, which, because Del~$S$ is a 
topological ball, result in $\X$ a ball and $\W$ a topological sphere.

To produce surfaces of higher genus, the contraction
is pushed through holes:
the most ``significant'' sink 
(in terms of $g(x)$)
is deleted
(changing the homotopy type), and then the collapses resume as before.
This is how the shapes shown in Figs.~1 and~2 were produced.
Repeating this process on the sorted sinks results in a series of
nested complexes $\X = \X_0, \X_1,\ldots, \emptyset$.

Finally, the algorithm works in any dimension, although most
applications are in $\R^3$.

\small
\bibliographystyle{/home1/orourke/tex/alphasort}
\bibliography{extra,/home1/orourke/bib/geom/geom}

\end{document}